\documentclass[10pt,twocolumn,letterpaper]{article}

\usepackage[pagenumbers]{wacv} % To force page numbers, e.g. for an arXiv version

\definecolor{wacvblue}{rgb}{0.21,0.49,0.74}
\usepackage[pagebackref,breaklinks,colorlinks,allcolors=wacvblue]{hyperref}
\usepackage[utf8]{inputenc} % allow utf-8 input
\usepackage[T1]{fontenc}    % use 8-bit T1 fonts
\usepackage{url}            % simple URL typesetting
\usepackage{booktabs}       % professional-quality tables
\usepackage{amsfonts}       % blackboard math symbols
\usepackage{nicefrac}       % compact symbols for 1/2, etc.
\usepackage{microtype}      % microtypography
\usepackage{graphicx}
\usepackage{pifont}
\usepackage{multirow} 
\usepackage{subcaption}
\usepackage{amsmath,amssymb}    % professional-quality tables
\usepackage[dvipsnames,table,xcdraw]{xcolor}
\usepackage{makecell}

\newcommand{\xmark}{\ding{55}}
\newcommand{\stageone}{\raisebox{0.7pt}{\textcircled{\raisebox{-0.85pt}{1}}}}
\newcommand{\stagetwo}{\raisebox{0.7pt}{\textcircled{\raisebox{-0.85pt}{2}}}}
\newcommand{\stagethree}{\raisebox{0.7pt}{\textcircled{\raisebox{-0.85pt}{3}}}}

\def\wacvPaperID{768} % *** Enter the WACV Paper ID here
\def\confName{WACV}
\def\confYear{2027}

\title{MADJEPA: Enhanced Multi-Stage Joint-Embedding Predictive Architecture for Medical Anomaly Detection}

\author{
    Ioana-Sabina Stoian\textsuperscript{\rm 1},
    Florinel-Alin Croitoru\textsuperscript{\rm 2},
    Vlad Hondru\textsuperscript{\rm 2},\\
    Andreea Iuliana Ionescu\textsuperscript{\rm 3,4},
    Andrei P\u{a}un\textsuperscript{\rm 1,2},
    Radu Tudor Ionescu\textsuperscript{\rm 2,$\diamond$}\\
    \textsuperscript{\rm 1}Romanian Academy Research Institute for Artificial
Intelligence, Romania\\
\!\textsuperscript{\rm 2}University of Bucharest, Romania\quad\!\textsuperscript{\rm 3}``Carol Davila'' University of Medicine and Pharmacy, Romania\\
 \textsuperscript{\rm 4}Col\c{t}ea Clinical Hospital, Romania\quad   \textsuperscript{ $\diamond$}Corresponding author: raducu.ionescu@gmail.com.
\vspace{-0.5cm}}

\begin{document}
\maketitle

\begin{abstract}
Anomaly detection in medical images is a complex task, requiring the development of a robust model using only healthy patient data. Without access to labeled data, modern approaches typically employ self-supervised learning to obtain a model of normality. Since world models promise to build strong internal representations of their environment, we consider them as suitable candidates for the medical anomaly detection environment, namely medical investigations of healthy patients. To this end, we propose MADJEPA, a \textbf{m}edical \textbf{a}nomaly \textbf{d}etector based on an enhanced multi-stage \textbf{j}oint-\textbf{e}mbedding \textbf{p}redictive \textbf{a}rchitecture (JEPA). We design a multi-stage training pipeline to produce a robust detector: (1) image-based JEPA (I-JEPA) training aims to learn a strong representation of normal training data; (2) target encoder distillation aims to modulate over-generalization of I-JEPA representations to abnormal data; (3) pseudo-anomaly classification aims to calibrate anomaly predictions by learning a tight boundary around the normal data manifold. To demonstrate our design, we carry out comprehensive experiments on six datasets covering various medical image domains (e.g.~MRI, OCT, X-ray, histopathology), comparing our MADJEPA with numerous state-of-the-art methods. The results show that our carefully designed world model, MADJEPA, attains competitive results, ranking first on three datasets. Our ablation studies further confirm the utility of each training stage, indicating that I-JEPA alone is not sufficient to obtain competitive results in medical anomaly detection. We publicly release our code at \small{\url{https://github.com/Sabina085/madjepa}}.
\end{abstract}

\setlength{\abovedisplayskip}{2.5pt}
\setlength{\belowdisplayskip}{2.5pt}
\setlength{\abovedisplayshortskip}{2.5pt}
\setlength{\belowdisplayshortskip}{2.5pt}

\vspace{-0.3cm}
\section{Introduction}
\label{sec:intro}
\vspace{-0.1cm}

Medical anomaly detection (MAD) aims to identify features that deviate from a highly diverse and partially characterized distribution of normal anatomy. The task poses additional challenges compared with the usual supervised recognition setting, where the objective is just to discriminate between a predefined set of pathological categories. In medical imaging, this open-set formulation \cite{Wang-Nature-2023,Bercea-MIDL-2024} is particularly relevant, as pathological samples are often rare and expensive to annotate \cite{Baur-MIA-2021,Kascenas-MIA-2023,Cai-MIA-2025}. 
In contrast, hospitals and imaging repositories contain large collections of unlabeled examinations, the vast majority of which exhibit normal anatomy or contain only weakly characterized findings \cite{Baur-MIA-2021,Bao-CVPRW-2024,Cai-MIA-2025}. These observations motivate unsupervised MAD: rather than attempting to enumerate every possible pathology, the model learns a representation of normal anatomy and identifies samples whose representations deviate from it \cite{Schlegl-IPMI-2017,Schlegl-MIA-2019,Baur-MIA-2021}.

The central problem is therefore to determine what constitutes a useful representation of normality. This question is more subtle in medical imaging than in natural-image anomaly detection. Normal anatomical structures exhibit substantial variation across patients, imaging modalities, acquisition protocols, anatomical regions, and disease prevalence. At the same time, abnormalities may manifest at very different levels of abstraction, \eg a lesion can differ from its surroundings through local texture, alter the spatial configuration of an organ, or disrupt anatomical context. Consequently, a representation that is sensitive to low-level appearance may miss contextual abnormalities, while an overly invariant representation may discard precisely the subtle local differences required for detection. The five medical domains from BMAD \cite{Bao-CVPRW-2024} provide a concrete illustration of how anomalies broadly vary, and demonstrate how methods that perform well in one domain do not necessarily transfer to another.

\begin{figure*}[t]
  \centering
   \includegraphics[width=0.76\linewidth]{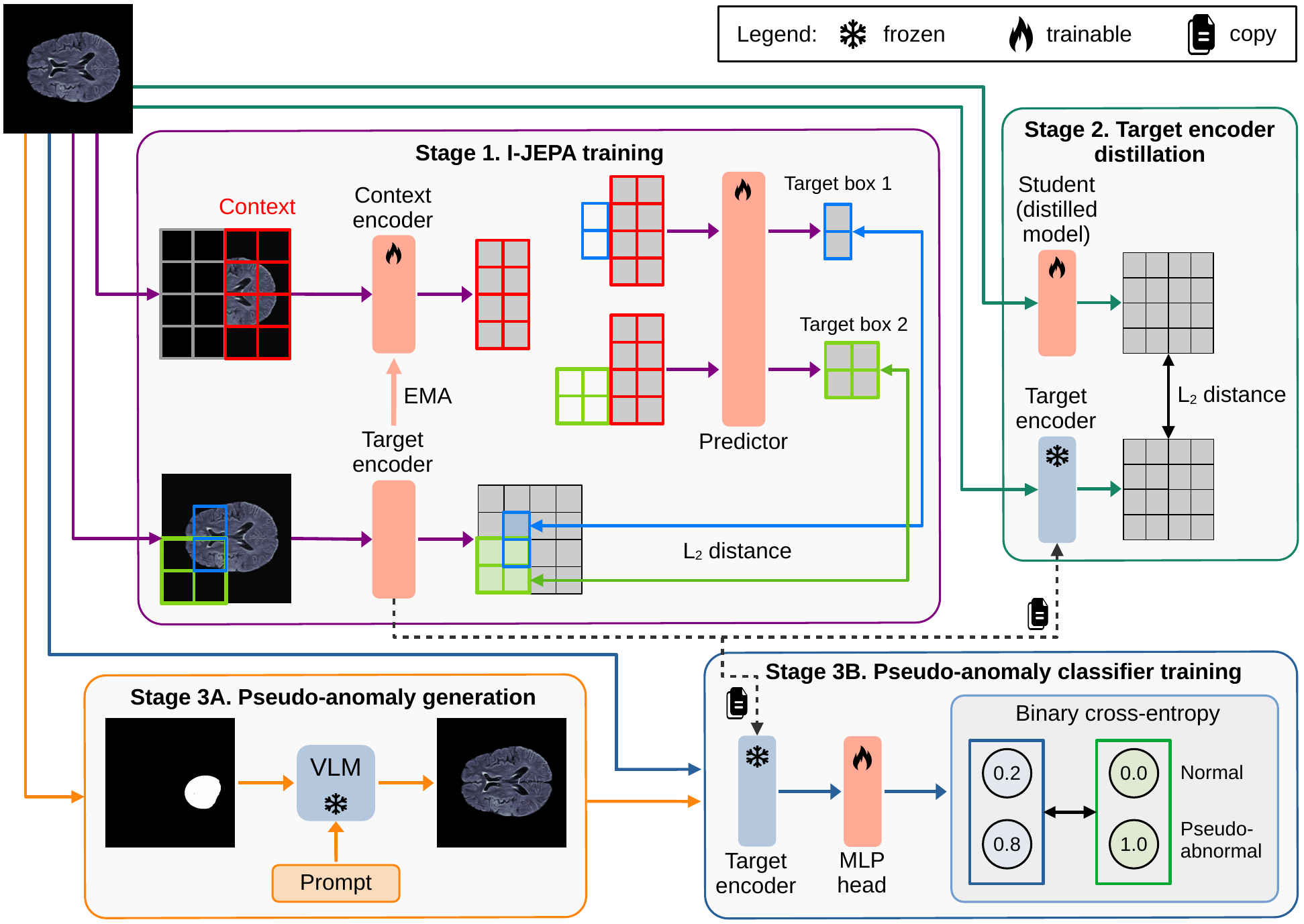}
   \vspace{-0.3cm}
   \caption{Overview of MADJEPA. Our pipeline is divided into three training stages: \stageone{} self-supervised training via I-JEPA; \stagetwo{} target encoder distillation to a randomly-initialized student; \stagethree{} A: prompt-based pseudo-anomaly generation, and B: normal vs.~pseudo-anomaly classification. Together, these training stages build a strong and calibrated representation of normal data. Best viewed in color.}
   \label{fig:pipeline_MADJEPA}
      \vspace{-0.4cm}
\end{figure*}

To address the above challenges, we tap into world models \cite{Assran-CVPR-2023,Balestriero-Arxiv-2025,Dawid-JSMTE-2024,Ha-NeurIPS-2018}, specifically I-JEPA (Image-based Joint-Embedding Predictive Architecture) \cite{Assran-CVPR-2023}, which aim to learn robust semantic representations instead of surface details learned by reconstructing raw input pixels. By predicting target features directly in an abstract latent space, a world model can filter out high-frequency noise, background variations, and irrelevant image artifacts, thus focusing on the true structural priors of healthy tissue. Because the network is trained as a predictive world model purely on normal anatomical structures, unseen anomalous patterns (\eg tumors, lesions, or out-of-distribution structures) break the learned latent space dependencies, generating an elevated latent prediction error that serves as a natural proxy for unsupervised MAD.

I-JEPA learns self-supervised representations via spatial context-block prediction without relying on handcrafted data augmentations (such as patch interpolation \cite {Tan-MLBI-2022}, color jittering \cite{Tian-MIA-2023} or synthetic intensity alterations \cite{Deng-EMBC-2022}) that can distort subtle pathological features or introduce artificial domain biases. While this increases the faithfulness of the representation for normal training data, it can lead to a poorly calibrated representation when normal training data is insufficient to comprehensively represent the full range of healthy anatomical structures. As a result, the model can either label unrecognized normal tissue as abnormal, or may over-generalize to represent subtle anomalies. To this end, we propose MADJEPA, a \textbf{m}edical \textbf{a}nomaly \textbf{d}etector based on an enhanced multi-stage \textbf{j}oint-\textbf{e}mbedding \textbf{p}redictive \textbf{a}rchitecture (JEPA). As illustrated in Figure \ref{fig:pipeline_MADJEPA}, we design a three-stage training pipeline to produce a robust medical anomaly detector. In stage \stageone, we perform self-supervised training via I-JEPA to learn a strong representation of normal training data. In stage \stagetwo, we distill the target encoder into a lighter randomly-initialized student, further harnessing the discrepancy between latent space predictions in stages \stageone{} and \stagetwo{} to modulate over-generalization of I-JEPA to abnormal data. In stage \stagethree, we generate pseudo-anomalies and learn to discriminate normal tissue from pseudo-anomalies, enabling the calibration of anomaly predictions by learning a tight boundary around the normal data manifold.

% We take a totally different perspective: learning semantic representations through latent feature prediction. This objective encourages the model to encode semantic context and structural relationships, producing representations that are less sensitive to pixel-level variations and potentially more effective for distinguishing normal from anomalous patterns.

We conduct comprehensive experiments on six datasets covering various medical image domains (e.g.~MRI, OCT, X-ray, histopathology), which form the recently proposed BMAD benchmark \cite{Bao-CVPRW-2024}. By comparing our MADJEPA with numerous state-of-the-art methods \cite{Schlegl-MIA-2019,Akcay-ACCV-2019,Zavrtanik-ICCV-2021,Chen-NN-2022,Ruff-ICML-2018,Li-CVPR-2021,Liu-CVPR-2023,Salehi-CVPR-2021,Deng-CVPR-2022,Yamada-arXiv-2022,Defard-ICPR-2021,Roth-CVPR-2022,Lee-IEEE-2022,Gudovskiy-WACV-2022,Rudolph-WACV-2022,Kim-PR-2024,dalmonte-wacv-2026}, we find that our carefully designed multi-stage world model, MADJEPA, attains competitive results, ranking first on three datasets, while being the only method surpassing the $80\%$ threshold on all six datasets. Furthermore, our ablation studies confirm the utility of each training stage, indicating that I-JEPA alone is not sufficient to obtain competitive results on the MAD task.

In summary, our contribution is threefold:
\begin{itemize}
    \item An enhanced multi-stage joint-embedding predictive architecture, where multiple training stages are integrated to obtain a robust and calibrated representation for MAD.
    \item Comprehensive experiments on six datasets comprising multiple medical imaging domains, where the proposed MADJEPA exhibits competitive results, often ranking first or second when compared with state-of-the-art methods.
    \item Ablation studies on the proposed training stages, confirming that each stage plays a significant role in boosting performance.
\end{itemize}

\vspace{-0.1cm}
\section{Related Work}
\label{sec:related_work}
\vspace{-0.1cm}

Prior work on medical anomaly detection can be grouped into four research directions: reconstruction-based methods that capture deviations from learned normal anatomy \cite{zimmerer-miccai-2019,zhou-tmi-2022,Baur-MIA-2021,cai-miccai-2024,sina-jmi-2024,pinaya-miccai-2022,behrendt-icmidl-2023,baugh-mlim-2025,yuan-mia-2025, beizaee-ipmi-2025}, foundation or vision-language models, some of which enable few-shot and zero-shot detection \cite{zhou-iclr-2024, huang-cvpr-2024, ma-cvpr-2025, jeong-cvpr-2023,zhu-cvpr-2024,gu-cvpr-2025}, unified learning frameworks that generalize across domains and modalities \cite{zhu-cvpr-2024,gu-cvpr-2025}, and multi-task learning strategies for building shared, generalizable representations \cite{jezequel-tip-2023, Tian-MIA-2023, behzad-miccai-2023}.

Medical anomaly detection has traditionally been dominated by reconstruction-based methods~\cite{zimmerer-miccai-2019,zhou-tmi-2022,Baur-MIA-2021,cai-miccai-2024,sina-jmi-2024}, which learn only from healthy samples and identify anomalies from reconstruction errors, \ie treating deviations in the reconstructed output as anomaly cues. More recently, leveraging diffusion models, which showcase strong generative capabilities \cite{Croitoru-TPAMI-2023}, the healthy counterparts of anomalous samples are synthesized, the difference between the two being analyzed to detect anomalous regions \cite{pinaya-miccai-2022,behrendt-icmidl-2023,baugh-mlim-2025,yuan-mia-2025,beizaee-ipmi-2025}. Most notably, Behrendt~\etal~\cite{behrendt-miccai-2024} improved the reconstruction robustness by estimating the anomaly likelihood from multiple  pseudo-healthy samples, then computing the Mahalanobis distance at the pixel-level. In order to preserve the healthy anatomical structure, Beizaee~\etal~\cite{beizaee-ipmi-2025} proposed masked latent diffusion to specifically carry the diffusion process only on the anomalous regions.

With the emergence of generic vision-language models (VLMs), particularly CLIP \cite{radford-ICML-2021}, researchers have started to employ such models for anomaly detection, even in few-shot and zero-shot regimes \cite{zhou-iclr-2024, huang-cvpr-2024, zhang-miccai-2024, shiri-miccai-2025, zhou-miccai-2025, ma-cvpr-2025, jeong-cvpr-2023}. Among these, Huang~\etal~\cite{huang-cvpr-2024} inserted residual adapters at multiple CLIP encoder stages and combined zero-shot text-based scoring with few-shot memory-bank comparison. Zhang~\etal~\cite{zhang-miccai-2024} proposed MediCLIP, fine-tuning CLIP on synthetic anomalous samples that mimic clinically plausible patterns, using adapted features from multiple levels and learnable prompt tokens, obtaining substantial gains using as much as $1\%$ of the training set. Shiri~\etal~\cite{shiri-miccai-2025} introduced MadCLIP, a dual-branch adapter trained contrastively with SigLIP loss to explicitly separate normal and abnormal feature subspaces. Ma~\etal~\cite{ma-cvpr-2025} proposed AA-CLIP, a two-stage sequential adaptation of both text and visual encoders. It builds anomaly-aware text anchors that capture distinct normal and abnormal semantics, before aligning patch-level image features with these to facilitate the localization of anomalies. These ideas have also been extended beyond binary anomaly detection to fine-grained disease classification, as demonstrated by UltraAD for ultrasound imaging~\cite{zhou-miccai-2025}.

More broadly, recent work has sought to improve cross-domain generalization by training unified models across heterogeneous anomaly detection benchmarks \cite{zhu-cvpr-2024,gu-cvpr-2025} or transferring learned anomaly representations to entirely unseen imaging modalities~\cite{mao-cvpr-2025}. Recently, Dalmonte~\etal~\cite{dalmonte-wacv-2026} demonstrated that a frozen vision foundation model combined with lightweight learnable bottlenecks can achieve strong unsupervised anomaly detection, and thus, the encoders do not require any training.

Another research direction relies on self-supervised learning, where models acquire semantic representations through proxy tasks without requiring pixel-level anomaly annotations. Existing methods combine objectives such as rotation prediction, jigsaw solving, recolorization \cite{jezequel-tip-2023}, contrastive learning, or synthetic anomaly generation~\cite{Tian-MIA-2023, behzad-miccai-2023}. However, these proxy tasks are typically optimized independently or combined only through score fusion at inference time, limiting their ability to learn complementary representations jointly. Similar to our method, Le~\etal~\cite{le-ACIIDS-2026} employed I-JEPA on medical images. However, their focus is on segmentation rather than detection, with a limited evaluation on chest X-rays. In contrast, we integrate I-JEPA in a multi-stage training pipeline, which brings significant gains over I-JEPA alone.

\vspace{-0.1cm}
\section{Method}
\label{sec:method}
\vspace{-0.1cm}

As illustrated in Figure~\ref{fig:pipeline_MADJEPA}, the proposed medical anomaly detector is based on an enhanced multi-stage joint-embedding predictive architecture. The training process is specifically divided into three stages, and the underlying architecture is based on a vision transformer (ViT) \cite{Dosovitskiy-ICLR-2021}. MADJEPA predicts both image-level and pixel-level anomaly scores based on discrepancies among latent representations.

\noindent\textbf{Architecture.}
In stage \stageone, we employ a ViT-S architecture for two of the three I-JEPA components, namely the context encoder and the target encoder, and a custom ViT-based architecture for the predictor. ViT consists of several transformer encoder layers, each comprising a multi-head self-attention module and a position-wise feed-forward network with GeLU activation, connected via skip-connections and pre-layer normalization. The context encoder and the target encoder employ Rotary Positional Embeddings (RoPE) \cite{su-Neurocomputing-2024}, while the predictor leverages sinusoidal positional embeddings \cite{vaswani-NeurIPS-2017}. The input image is resized to a fixed resolution $H \times W$, and partitioned into $K = (H / P) \times (W / P)$ non-overlapping patches, each patch being projected into a $d$-dimensional vector via a patch encoder, implemented as a 2D convolutional layer. 
% All three models are pretrained on \td{ImageNet? using DINOv3 \cite{}.} large scale datasets of natural images. 
Both the target encoder and the context encoder are pretrained on the LVD-1689M \cite{simeoni-arxiv-2025} dataset using DINOv3 \cite{simeoni-arxiv-2025}. The context encoder shares the same architectural configuration as the target encoder and is initialized with the same weights. The predictor model is pretrained on the ImageNet-22K dataset following the I-JEPA framework \cite{Assran-CVPR-2023}.
For the student model introduced in stage \stagetwo{}, we employ a lighter variant of the ViT architecture, comprising a reduced number of transformer encoder layers, with randomly initialized weights. In stage \stagethree{}, the pseudo-anomaly classifier leverages an MLP head consisting of two dense layers with GeLU activation.

\noindent\textbf{Stage 1: I-JEPA training.}
Many self-supervised methods learn image representations by leveraging handcrafted augmentations. However, such artificial perturbations may be detrimental to the unsupervised anomaly detection task, as they can enable the model to be less sensitive to small abnormal shifts, ultimately resulting in a failure to detect subtle anomalies. At the same time, an effective method must remain robust to slight variations within the normal space, requiring its representations to encode semantic context and structural relationships from the normal data. I-JEPA \cite{Assran-CVPR-2023} represents a strong candidate in this regard, as it addresses both limitations. More specifically, without relying on synthetic augmentations, I-JEPA learns semantically rich representations through latent space feature prediction. By constructing prediction targets directly in the latent space, I-JEPA learns representations that are highly semantic and less sensitive to pixel-level variations, making them potentially more effective for distinguishing normal from abnormal patterns. To this end, our first training stage is based on I-JEPA \cite{Assran-CVPR-2023}. A predictor $\mathbf{P}$ is trained to reconstruct the target encoder's latent space representations of masked image regions, relying only on the encoding of a limited visible context. First, the target encoder $\mathbf{T}$ processes the complete (unmasked) image, generating representations in the latent space for each patch, denoted as $\mathcal{S}_\mathbf{T} = \{\mathbf{t}_i\}_{i=\overline{1, K}}$, where $\mathbf{t}_i \in \mathbb{R}^d$ is the representation of the patch at index $i$. Let $\mathcal{B} = \bigcup_{j=1}^{M} {\mathcal{B}_j}$ be the union of patch index sets of the $M$ randomly sampled target blocks. Second, a partially masked image is passed to the context encoder in order to generate rich semantic representations for the patches contained in the visible context. Subsequently, the predictor takes as input the encoded context and a mask token for each patch that needs to be predicted, and outputs per-patch representations $\mathcal{S}_\mathbf{P} =\{\mathbf{p}_i\}_{i \in \mathcal{B}}$, where $\mathbf{p}_i \in \mathbb{R}^d$ is the predictor representation of the patch at index $i$.
The training objective enforces similarity between the latent representations generated by the predictor ($\{\mathbf{p}_i\}_{i \in \mathcal{B}_j}$) and the target encoder ($\{\mathbf{t}_i\}_{i \in \mathcal{B}_j}$) for the $M$ randomly sampled target blocks $\{\mathcal{B}_j\}_{j=\overline{1, M}} $. To this end, we employ a smooth $L_1$ loss \cite{girshick-ICCV-2015} between the two representations as the objective function. For an input image $\mathbf{X}$, the objective is formally defined as:
\begin{equation}
    \label{eq:loss-ijepa}
    \mathcal{L}_\text{JEPA}(\mathbf{X}) = \frac{1}{M  \cdot |\mathcal{B}| \cdot d}\sum_{j=1}^{M}\sum_{i \in \mathcal{B}_j}\sum_{k=1}^{d} l_{i, k}, \text{with:} 
\end{equation}
\begin{equation}
l_{i, k} =
\begin{cases}
    \label{eq:li_loss}
    0.5 \cdot (\mathbf{p}_{i, k} - {\mathbf{t}_{i, k}})^2, & | {\mathbf{p}_{i, k}} - {\mathbf{t}_{i, k}}| < 1 \\
    | {\mathbf{p}_{i, k}} - {\mathbf{t}_{i, k}}| - 0.5, & \text{otherwise}.
\end{cases}
\end{equation}
We train the entire system via full fine-tuning. The context encoder and the predictor are optimized via gradient descent, while the target encoder is updated using an exponential moving average (EMA) of the context encoder weights.

\noindent\textbf{Stage 2: Target encoder distillation.}
This stage leverages a lightweight ViT student which is trained to perform knowledge distillation (KD) from the I-JEPA target encoder obtained in stage \stageone{}. The integration of a distilled target encoder model into our anomaly detection pipeline is primarily motivated by the need to mitigate over-generalization. Since the target encoder was pretrained on a large-scale dataset with natural images and subsequently fine-tuned on medical data (via stage \stageone{}), it may exhibit a tendency to over-generalize on anomalous regions. To address this limitation, we introduce a student model, a lightweight ViT trained from scratch, to replicate the output of the target encoder on normal training images. Given that the student's weights are randomly initialized and its training is restricted exclusively to normal samples, it possesses no prior knowledge of how anomalies are represented in the latent space. We hypothesize that, when presented with an anomalous region, the distilled student model will diverge in behavior from the more capable target encoder. This behavioral divergence is expected to manifest as an increased discrepancy in the latent space representations, thereby serving as a reliable indicator of potentially anomalous regions.

The student model $\mathbf{S}$ is trained to replicate the latent representations of normal samples generated by the teacher (target encoder) $\mathbf{T}$. Both models receive an unmasked image as input and produce, for each patch, an embedding vector that aggregates the semantic information contained within the corresponding image region. The objective of the student is to produce patch-level latent representations $\mathcal{S}_\mathbf{S} = \{\mathbf{s}_i\}_{i=\overline{1, K}}$ that closely approximate those of the teacher \textbf{$\mathcal{S}_\mathbf{T} = \{\mathbf{t}_i\}_{i=\overline{1, K}}$}, where $\mathbf{s}_i \in \mathbb{R}^d$ is the student representation of the patch at index $i$, and $\mathbf{t}_i \in \mathbb{R}^d$ is defined in stage \stageone{}. To this end, we perform target encoder distillation using a mean squared error objective:
\begin{align}
    \label{eq:loss-distill}
    \mathcal{L}_{\text{KD}} (\mathbf{X}) = \frac{1}{K}\sum_{i=1}^{K} \| \mathbf{s}_i - \mathbf{t}_i \|^{2}_{2}.
\end{align}
% where $\mathbf{X}$ represents an original (unmasked) image. 

\noindent\textbf{Stage 3: Pseudo-anomaly generation and classification.}
This stage comprises two sequential phases. In phase A, pseudo-abnormal images are generated from the normal training images. In phase B, a classification head is trained on top of the frozen I-JEPA target encoder backbone to distinguish between normal and pseudo-abnormal samples. 

\noindent {\color{ForestGreen}\emph{Phase A:}}
We generate pseudo-abnormal samples by leveraging Qwen-Image~\cite{Wu-arXiv-2025}. Specifically, we modify a healthy sample using a prompt and a mask, via inpainting. The prompts are specific to the medical domain of the real image that is perturbed. We generate the inpainting masks using a random number of overlapping ellipses of varying sizes and locations. To ensure the inpainting process targets actual tissue in images with extensive background regions, we constrain the ellipse centers within organ boundaries. Ultimately, this randomized overlapping approach naturally yields final masks with highly diverse and complex shapes. Examples of generated images are in the supplementary.

\noindent {\color{ForestGreen}\emph{Phase B:}} 
We train a classifier to discriminate between normal samples and pseudo-abnormal samples synthesized in phase A. The introduction of this classification stage is motivated by the fact that the appearance of anomalous regions is highly specific to both the medical domain and the acquisition modality, while the boundary around the normal data manifold remains too loose, and may fail to capture subtle anomalies that require a fine distinction from normal samples. To tighten the detection boundary around the normal data manifold, we introduce a separate prediction task in which a binary classifier is trained to separate normal and pseudo-abnormal samples.

% the visual inspection of the validation data, which revealed that the appearance of anomalous regions is highly specific to both the medical domain and the acquisition modality. For instance, in the Brain MRI dataset, anomalous regions predominantly appear as hyperintense (light-pixel) areas, possibly related to the administration of a contrast substance, while darker regions are more consistent with normal tissue. Conversely, in the Liver CT dataset, lesions tend to appear as hypointense (dark-pixel) blobs that can be easily confounded with the background, whereas lighter regions are more often associated with normal tissue. 

We employ the target encoder trained in stage \stageone{} as a frozen feature extractor and train a classifier on top of its pooled latent representation. Specifically, the patch-level representations $\mathcal{S}_\mathbf{T} = \{\mathbf{t}_i\}_{i=\overline{1, K}}$ produced by the target encoder are aggregated via mean pooling into a compact image embedding vector: $\hat{\mathbf{t}} = \frac{1}{K}\sum_{i=1}^{K} \mathbf{t}_i \in \mathbb{R}^d$. This global representation is passed as input to an MLP classifier consisting of two fully-connected layers with intermediate GeLU activation and a final sigmoid non-linearity: $\hat{y} = \sigma\left(\text{MLP}(\hat{\mathbf{t}})\right)$.
The classification head predicts a probability that indicates if a given input image contains a synthetic anomaly. The model is trained on both normal and pseudo-abnormal images using the binary cross-entropy loss:
\begin{align}
    \label{eq:loss-cls}
    \mathcal{L}_{\text{cls}} (\mathbf{X},\ y) = - (y \cdot \log \hat{y} + (1 - y) \cdot \log(1 - \hat{y})).
\end{align}

\noindent\textbf{Inference.}
Our anomaly detection pipeline builds upon the trained system described in the three preceding stages, whose outputs are combined to produce the final image-level and pixel-level anomaly scores. 

% TOO DETAILED
In stage \stageone{}, a context block and a target block are randomly sampled, using the sampling procedure employed during training. The entire target block is masked (including the overlapping region between context and target blocks), and the remaining visible patches are encoded via the context encoder. Subsequently, the predictor processes the sequence of context embeddings and produces latent representations for the masked target patches. Concurrently, the target encoder processes the full image to generate embedding vectors for all constituent patches. For each patch within the sampled target block, a distance metric (\eg $L_1$ or $L_2$) is computed between the corresponding latent representations produced by the predictor and the target encoder, respectively. This procedure is repeated across $N$ iterations with a new context block and a different target block resampled at each step. As a result, multiple distance values are obtained per patch. %, which are subsequently aggregated using a pooling function (\eg mean or max). 
The resulting per-patch values are aggregated into an anomaly map $A_{\text{JEPA}}$ of size $(H / P) \times (W / P)$. The map is resized to the input image resolution using bilinear interpolation, smoothed via a Gaussian blurring filter, and scaled to the $[0, 1]$ range. Regions exhibiting larger aggregated distances will appear as high-intensity pixel groups, suggesting a higher likelihood of anomalous content, whereas regions with smaller distances are indicative of normal tissue. The image-level AUC score, in turn, is obtained by aggregating the values of the anomaly map via a pooling function (\eg~max or mean), producing a single anomaly score per image, denoted as $a_{\text{JEPA}}$.

In stage \stagetwo{}, the input image is provided to both the teacher and the student models, each producing an embedding vector for each patch. A distance metric (\eg $L_1$ or $L_2$) is then computed between the corresponding patch-level representations of the two models. Following the same procedure outlined in stage \stageone{}, the resulting per-patch distance values are used to construct a second anomaly map $A_{\text{KD}}$, from which an image-level anomaly score $a_{\text{KD}}$ is derived.

In stage \stagethree{}, the full image is passed to the pseudo-anomaly classifier, which outputs the probability that the image represents a synthetic anomaly sample (denoted $a_{\text{cls}}$). 

The three image-level scores are subsequently combined via a weighted mean using coefficients $\lambda_1$, $\lambda_2$ and $\lambda_3$ in order to produce the final anomaly score:
\begin{equation}
    \label{eq:anomaly_maps_weighting_image}
    a_{\text{final}} = \lambda_1 \cdot a_{\text{JEPA}} + \lambda_2 \cdot a_{\text{KD}} + \lambda_3 \cdot a_{\text{cls}},
\end{equation}
where $\sum_{i=1}^{3}\lambda_i = 1$, $\lambda_i \in [0, 1]$.

%For datasets with pixel-level annotations (namely BraTS2021\cite{Baid-arXiv-2021}, BTCV\cite{Landman-MICCAI-2015} + LiTS \cite{Bilic-MIA-2023} and RESC\cite{Hu-MIA-2019}), the generated anomaly maps are directly employed to compute the pixel-level AUC score, using the available segmentation masks as ground truth. To this end, 

Similarly, the anomaly maps generated in stage \stageone{} and \stagetwo{} are aggregated through a weighted combination. The corresponding coefficients ($\lambda_1$ and $\lambda_2$), introduced previously, are normalized by their sum prior to aggregation, ensuring that the resulting weights sum to $1$:
\begin{align}
    \label{eq:anomaly_maps_weighting_pixel}
    A_{\text{final}} & = \frac{\lambda_1}{\lambda_1 + \lambda_2} \cdot A_{\text{JEPA}} + \frac{\lambda_2}{\lambda_1 + \lambda_2} \cdot A_{\text{KD}}.
\end{align}
%where $AM_{ijepa}, AM_{distill} \in \mathbb{R^{H \times W}}$ are the resized anomaly maps from stage \stageone{} and stage \stagetwo{} respectively.

\vspace{-0.1cm}
\section{Experiments}
\label{sec:experiments}
\vspace{-0.1cm}

\begin{table*}[t]
    \centering
    \setlength\tabcolsep{0.05cm}
    %\resizebox{0.99\linewidth}{!}
    \fontsize{7.5}{8}\selectfont{
    \begin{tabular}{|l | cc | cc | cc | c | c | c|}
        \hline
        \multirow{2}{*}{Method} & \multicolumn{2}{c|}{BraTS2021} & \multicolumn{2}{c|}{BTCV+LiTS} & \multicolumn{2}{c|}{RESC} & {OCT2017} & {RSNA} & {CAMELYON16} \\
        %\cline{2-10}
         & Image AUROC & Pixel AUROC & Image AUROC & Pixel AUROC &  Image AUROC & Pixel AUROC & Image AUROC & Image AUROC & Image AUROC   \\ 
        \hline
        \hline
         f-AnoGAN~\cite{Schlegl-MIA-2019} & $77.3 \pm 0.18$ & NA & $58.5 \pm 0.21$ & NA & $77.4 \pm 0.85$ & NA & $73.4 \pm 1.85$ & $55.2 \pm 0.09$ & $69.5 \pm 1.98$ \\
         GANomaly~\cite{Akcay-ACCV-2019}& $74.8 \pm 1.93$ & NA & $54.6 \pm 3.06$ & NA & $52.6 \pm 3.95$ & NA & $70.5\pm 9.98$ & $62.9 \pm 0.65$ &  $54.4 \pm 2.57$ \\
         DRAEM~\cite{Zavrtanik-ICCV-2021} & $62.4 \pm 9.03$ & $82.3 \pm 4.07$ & $70.0 \pm 3.86$ & $87.5 \pm 3.23$ & $83.2 \pm 8.21$ & $86.8 \pm 3.14$ & $88.0 \pm 8.36$ & $67.7 \pm 1.72$ & $52.4 \pm 0.77$ \\
         UTRAD~\cite{Chen-NN-2022}  & $82.9 \pm 2.32$ & $ 92.6 \pm 0.67$ & $55.8 \pm 5.66$ & $ 87.9 \pm 1.32$ & $89.4 \pm 1.92$ & $94.5 \pm 1.24$ & $96.8 \pm 0.56$ & $75.6 \pm 1.24$ & $70.0 \pm 4.64$ \\
         % \\ 
         % \hline
         % \hline
         % \\
         
        DeepSVDD~\cite{Ruff-ICML-2018}  & $87.0 \pm 0.66$ & NA & $54.0 \pm 1.84$ & NA & $74.1 \pm 1.29$ & NA & $76.8 \pm 1.37$ & $64.5 \pm 3.17$ & $61.0 \pm 1.82$ \\
        
        CutPaste~\cite{Li-CVPR-2021}  & $78.8 \pm 0.67$ & NA & $59.3 \pm 4.86$ & NA & $90.2 \pm 0.61$ & NA & $96.8 \pm 0.62$ & $82.6 \pm 1.22$ & $75.2 \pm 0.41$ \\
        
        SimpleNet~\cite{Liu-CVPR-2023} & $82.5 \pm 3.34$ & $94.8 \pm 1.04$ & {\color{orange}$\underline{72.3 \pm 2.68}$} & {\color{orange}$\underline{97.5 \pm 0.56}$} & $76.2 \pm 7.46$ & $77.1 \pm 4.76$ & $94.7 \pm 2.17 $ & $69.1 \pm 1.27$ & $62.4 \pm 3.71$ \\
             
        MKD~\cite{Salehi-CVPR-2021} & $81.5 \pm 0.36$ & $89.4 \pm 0.24$ & $60.7 \pm 1.19$ & $96.1 \pm 0.27$ & $89.0 \pm 0.25$ & $86.7 \pm 0.65$ & $96.7 \pm 0.26$ & $82.0 \pm 0.12$ & \color{orange}\underline{$77.5 \pm 0.27$} \\
             
        RD4AD~\cite{Deng-CVPR-2022} & $89.5 \pm 0.91$ & $96.5 \pm 0.17$ & $60.4 \pm 1.17$ & $96.0 \pm 1.19$ & $87.8 \pm 0.87$ & \color{orange}$\underline{96.2 \pm 0.15}$ & $97.3 \pm 0.79$ & $67.6 \pm 1.11$ & $66.8 \pm 0.71$ \\
            
        STFPM~\cite{Yamada-arXiv-2022}  & $83.0 \pm 0.67$ & $95.6 \pm 0.12$ & $61.8 \pm 1.58$ & $91.2 \pm 5.52$ & $84.8 \pm 0.50$ & $94.7 \pm 0.57$ & $96.8 \pm 0.23$ & $72.9 \pm 1.96$ & $66.4 \pm 1.01$ \\
        
        PaDiM ~\cite{Defard-ICPR-2021} & $79.0 \pm 0.38$ & $94.4 \pm 1.03$ & $50.8 \pm 0.61$ & $90.9 \pm 0.84$ & $75.9 \pm 0.54$ & $91.4 \pm 0.42$ & $91.8 \pm 0.96$ & $77.5 \pm 1.87$ & $67.3 \pm 0.32$ \\
        
        PatchCore~\cite{Roth-CVPR-2022}  & $91.7 \pm 0.36$ & $97.0 \pm 0.04$ & $60.3 \pm 0.76$ & $ 96.4 \pm 0.19$ & \color{orange}\underline{$91.6 \pm 0.10$} & \color{blue}$\bold{96.5 \pm 0.10}$ & \color{orange}\underline{$98.6 \pm 0.03$} & $76.1 \pm 0.67$ & $69.3 \pm 0.21$ \\
        
        CFA~\cite{Lee-IEEE-2022}  & $84.4 \pm 0.87$ & $96.3 \pm 0.14$ & $62.0 \pm 1.08$ & $97.2 \pm 0.14$ & $69.9 \pm 0.26$ & $91.1 \pm 0.87$ & $79.5 \pm 0.56$ & $66.8 \pm 0.23$ & $65.6 \pm 0.10$ \\
        
        CFLOW~\cite{Gudovskiy-WACV-2022} & $74.8 \pm 5.32$ & $93.8 \pm 0.67$ & $50.8 \pm 4.47$ & $92.4 \pm 1.16$ & $75.0 \pm 5.81$ & $ 93.8 \pm 0.57$ & $85.4 \pm 2.11$ & $71.5 \pm 1.49$ & $55.7 \pm 1.97$ \\
        
        CS-Flow~\cite{Rudolph-WACV-2022}  & $90.9 \pm 0.83$ & NA & $59.4 \pm 0.54$ & NA & $87.3 \pm 0.58$ & NA & $98.5 \pm 0.28$ & $83.2 \pm 0.46$ & $68.4 \pm 0.42$ \\ 
        
        P-VQ~\cite{Kim-PR-2024}  &  {\color{blue}$\bold{94.3 \pm 0.23}$} & \color{blue}$\bold{98.4 \pm 0.02}$ & $60.6 \pm 0.62$ & $96.4 \pm 0.06$ & $89.0 \pm 0.48$ & $92.8 \pm 0.35$ & \color{blue}$\bold{99.3 \pm 0.22}$ & $79.2 \pm 0.04$ & NA \\ 
        
        QFAE~\cite{dalmonte-wacv-2026}  &  {\color{blue}$\bold{94.3 \pm 0.18}$} & NA & $65.5 \pm 1.96$ & NA & {\color{blue}$\bold{91.8 \pm 0.55}$} & NA & NA & {\color{orange}$\underline{83.8 \pm 0.46}$} & NA \\
        \hline
        MADJEPA (Ours) & {\color{orange}$\underline{94.1 \pm 0.06}$} & \color{orange}$\underline{97.3 \pm 0.07}$ & {\color{blue}$\bold{84.1 \pm 0.20}$} & {\color{blue}$\bold{99.5 \pm 0.15}$} & $88.4 \pm 0.16$ & $94.9 \pm 0.22$ & $96.9 \pm 0.26$ & {\color{blue}$\bold{84.4 \pm 0.28}$} & {\color{blue}$\bold{82.0 \pm 0.01}$} \\
    \hline
    \end{tabular}}
    \vspace{-0.25cm}
    \caption{Anomaly detection performance of our method MADJEPA on the complete BMAD benchmark \cite{Bao-CVPRW-2024}. We compare our method against a comprehensive set of state-of-the-art anomaly detection techniques from the literature, achieving the best performance on three challenging datasets (BTCV+LiTS, RSNA, CAMELYON16), and second-best on one dataset (BraTS2021). MADJEPA is the only method surpassing the $80\%$ threshold on all datasets and metrics. The best score on each dataset is highlighted in {\color{blue}\textbf{blue bold}} and the second-best score is highlighted in {\color{orange}\underline{orange underline}}. We report mean image-level and pixel-level AUROC scores (where pixel-level annotations are available), along with the corresponding standard deviations, computed across five different seeds. Best viewed in color.} 
    \label{tab:main_results}
    \vspace{-0.35cm}
\end{table*}

\noindent
\textbf{Datasets.}
We evaluate MADJEPA on all six datasets from BMAD \cite{Bao-CVPRW-2024}, a comprehensive benchmark for medical anomaly detection. These datasets cover five distinct medical domains: brain magnetic resonance imaging (BraTS2021~\cite{Baid-arXiv-2021}), liver computed tomography (BTCV~\cite{Landman-MICCAI-2015} + LiTS~\cite{Bilic-MIA-2023}), retinal optical coherence tomography (RESC~\cite{Hu-MIA-2019}, OCT2017~\cite{Kermany-Cell-2018}), Chest X-ray (RSNA~\cite{Wang-CVPR-2017}) and digital histopathology (CAMELYON16~\cite{Ehteshami-JAMA-2017}).
Pixel-level annotations are available for three of the six datasets (BraTS2021, BTCV+LiTS and RESC), whereas image-level annotations are provided across all six. More details about the six datasets are provided in supplementary. In stage \stagethree{}, a pseudo-abnormal image is generated for each normal sample in the training set, doubling the amount of training data.

\noindent
\textbf{Architecture details.}
In stage \stageone{}, the context and target encoders are DINOv3 \cite{simeoni-arxiv-2025} ViT-S/16 backbones with $12$ transformer layers, $6$ attention heads, and a hidden size of $384$. The predictor uses a separate pretrained ViT with $16$ layers, $16$ heads, and a hidden size of $384$. Following the official I-JEPA implementation~\cite{Assran-CVPR-2023}, each image is sampled with one context block covering $85$-$100\%$ of the image, and four target blocks with a scale of $15$-$20\%$. In stage \stagetwo{}, the student is a lightweight ViT-S/16 with $6$ transformer layers, $6$ attention heads, and a hidden size of $384$. Its weights are randomly initialized, with the target encoder copied from stage \stageone{} and kept frozen. In stage \stagethree{}, patch-level representations from the frozen target encoder are mean-pooled and projected to a $128$-dimensional representation using a linear layer with GeLU activation, followed by a second linear layer for the final pseudo-anomaly recognition.

\noindent
\textbf{Evaluation measures.} We report the standard evaluation metrics for BMAD, namely the image-level and pixel-level AUROC values.

\noindent
\textbf{Hyperparameters.}
For all training stages, we tune the hyperparameters of MADJEPA on the corresponding official validation splits. All baselines receive similar attention regarding hyperparameter tuning. %reported in Supplementary. 

In our experiments, we use a patch size of 16, and resize all images to $224 \times 224$ pixels. For stage \stageone{}, we follow the training procedure described in I-JEPA~\cite{Assran-CVPR-2023}, sampling four target blocks per input image and subsequently removing from the context all patches contained within the sampled boxes. The model is fully fine-tuned for $20$-$60$ epochs using AdamW with linear warm-up followed by cosine annealing. The learning rate is warmed up for $3$-$10$ epochs from $5\times10^{-5}$ to $10^{-3}$, then decayed to $10^{-6}$. We use a batch size of $32$, random cropping with minimum crop scales of $0.3$-$0.8$, and a target-encoder momentum schedule from $0.996$ to $1.0$. For stage \stagetwo{}, the student is trained for $40$-$100$ epochs with the same optimizer and scheduling strategy, using learning rates from $10^{-6}$ to a peak of $10^{-3}$ and a final value of $10^{-6}$. Batch sizes range from $64$ to $512$ and weight decay from $0.01$ to $0.05$. In stage \stagethree{}, only the classifier is optimized with AdamW using learning rates between $10^{-4}$ and $3\times10^{-3}$. The learning rate is halved after three epochs without validation AUROC improvement.

As far as the inference hyperparameters are concerned, we set $N=30$ across all datasets, and compare mean and maximum aggregation at both patch and image levels. $L_1$ and $L_2$ distances are evaluated for patch-level discrepancies. All other hyperparameters, such as the task weights, $\lambda_1$, $\lambda_2$, and $\lambda_3$, as well as the Gaussian smoothing parameters, are specified in the supplementary.

% An ablation on $N$ is provided in Section~\ref{sec:ablations}.

%The distance metric ($L_2$ or $L_1$), aggregation functions (max or mean), number of sampled target blocks and the weighting coefficients $\lambda_1$, $\lambda_2$ and $\lambda_3$  were selected independently per dataset via empirical validation on the corresponding official validation splits.

\begin{figure*}[t]
  \centering
   \includegraphics[width=0.8\linewidth]{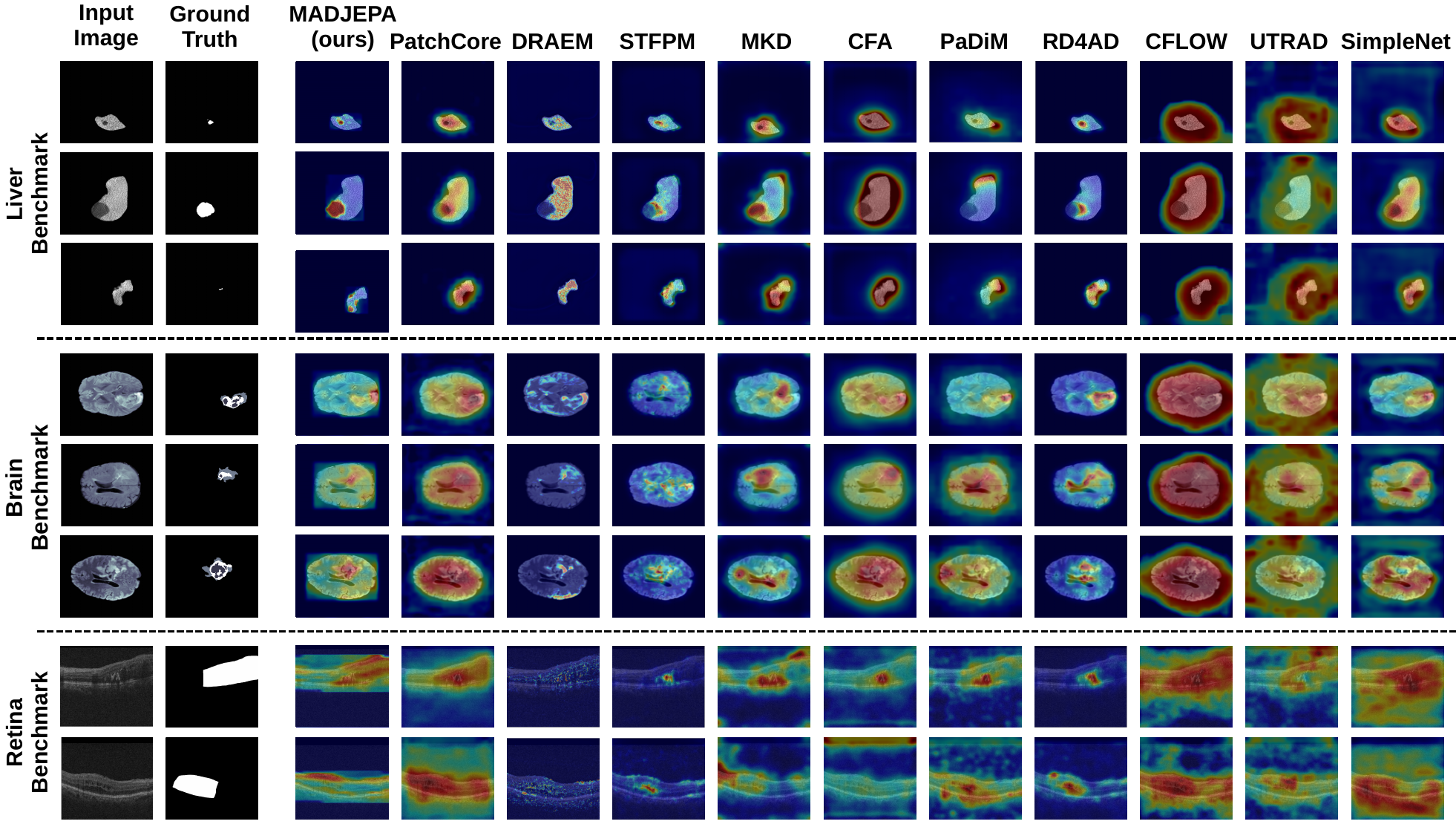}
   \vspace{-0.35cm}
   \caption{Qualitative visualizations of anomaly maps for the three benchmarks that have ground-truth segmentation maps: BTCV+LiTs, BraTS2021 and RESC. We employ the exact examples selected by Bao et al.~\cite{Bao-CVPRW-2024} from the BMAD original paper. Best viewed in color.}
   \label{fig:anomaly_localization}
   \vspace{-0.35cm}
\end{figure*}

\noindent
\textbf{Main results.} In Table ~\ref{tab:main_results}, we compare MADJEPA against a wide range of state-of-the-art anomaly detection methods on the complete BMAD benchmark. For each experiment, we report the mean and standard deviation of performance across five runs, varying the seed. MADJEPA achieves state-of-the-art performance on three datasets (BTCV+LiTS, RSNA and CAMELYON16), the second-best score on BraTS2021, and competitive results on the two retinal OCT datasets, RESC and OCT2017, where it outperforms the majority of baselines.

On the most challenging dataset, namely liver CT (BTCV+LiTS), our method substantially surpasses ($+11.8\%$) the previous state-of-the-art established by SimpleNet \cite{Liu-CVPR-2023}, achieving an image-level AUROC of $84.1 \pm 0.20$. %This represents an improvement of over $11\%$ in AUROC relative to the next-best score of $72.3 \pm 2.68$. 
We observe another significant improvement for MADJEPA in the second-most difficult dataset (CAMELYON16), where it attains an image-level AUROC of $82.0 \pm 0.01$, surpassing MKD \cite{Salehi-CVPR-2021} ($77.5 \pm 0.27$), the strongest baseline, by $4.5\%$. We further note that two recent methods, P-VQ \cite{Kim-PR-2024} and QFAE \cite{dalmonte-wacv-2026}, are particularly good on certain datasets, \eg BraTS2021, while lagging far behind on others, \eg BTCV+LiTS, and even missing results on some challenging datasets, \eg CAMELYON16. Although MADJEPA is not always a top-performer, \emph{its results are more consistent across the board, being the only method surpassing the $80\%$ threshold on all six datasets, in terms of both image-level and pixel-level AUROC.}

% On the chest X-ray dataset (RSNA), MADJEPA ($84.4 \pm 0.28$) exceeds the next-best result of $83.8 \pm 0.46$ obtained by QFAE \cite{dalmonte-wacv-2026}, corresponding to a gain of $0.6\%$ in terms of AUROC. On the brain MRI dataset (BraTS2021), MADJEPA achieves an image-level AUROC of $94.1 \pm 0.06$, closely approaching the current state-of-the-art achieved by P-VQ \cite{Kim-PR-2024} ($94.3 \pm 0.23$) and QFAE \cite{dalmonte-wacv-2026} ($94.3 \pm 0.18$). 

\noindent
\textbf{Qualitative results.} At the pixel level, the relative ranking of our method is consistent with the trends observed at the image level. The strong results prove the effectiveness and versatility of MADJEPA in detecting anomalies across diverse medical imaging datasets. In Figure ~\ref{fig:anomaly_localization}, we present some estimated anomaly maps using the proposed method. For visualization purposes, the anomaly maps are displayed as heatmaps, with high-intensity red regions denoting elevated anomaly likelihood and dark blue regions indicating normal tissue. 

On the BTCV+LiTS dataset, DRAEM and RD4AD tend to underestimate lesion extent, producing smaller and more restrictive responses, whereas CFA, CFLOW, and UTRAD tend to overestimate it, generating activations that extend into the surrounding normal hepatic parenchyma. PatchCore, STFPM, MKD, PaDiM, and SimpleNet show an intermediate behavior. However, in the third liver example, MADJEPA correctly identifies the lesion, but also generates two additional foci of activation with no corresponding ground-truth abnormalities, representing false-positive responses. Even in the presence of these isolated false-positive activations, MADJEPA provides, overall, the most balanced visual correspondence with the ground truth, avoiding both the systematic underestimation and the marked spatial overextension observed with the competing methods.
On the BraTS2021 dataset, DRAEM and RD4AD tend to underestimate the pathological region, whereas CFA, CFLOW, and UTRAD show the most evident overestimation, with anomaly responses extending into brain regions without corresponding pathological findings. PatchCore, STFPM, MKD, PaDiM, and SimpleNet exhibit varying degrees of deviation from the ground truth. Compared with these methods, MADJEPA shows better spatial correspondence with the ground truth, concentrating its activation within the pathological regions, while limiting nonspecific extension into the surrounding normal brain parenchyma.
On the RESC dataset, DRAEM and RD4AD again generate the most restrictive responses, suggesting underestimation of anomaly extent, whereas CFLOW, UTRAD, and SimpleNet produce more extensive activations, consistent with overestimation. PatchCore, STFPM, MKD, CFA, and PaDiM show an intermediate behavior. MADJEPA does not perfectly reproduce the ground-truth boundaries at the retinal level. However, the regions of highest activation remain predominantly associated with the pathological changes. 

Overall, MADJEPA provides a favorable visual balance between the two major error patterns observed among competing methods (underestimation of lesion extent and overestimation through activation of normal tissue), while maintaining this behavior across three distinct anatomical and imaging contexts.

%They were obtained by aggregating the stage \stageone{} (I-JEPA module) and stage \stagetwo{} (target encoder distillation) anomaly maps through a weighted mean, as formalized in Equation~\ref{eq:anomaly_maps_weighting_pixel}. This aggregated map was then upsampled to the input image resolution via bilinear interpolation, followed by Gaussian smoothing and scaled to the [0, 1] range. As it can be observed, our method manages to capture the anomalous regions very well, having an especially high fidelity on the liver benchmark.

\begin{table}[t!]
    \centering
    \fontsize{7.5}{8}\selectfont{
        \begin{tabular}{|l|ccc|cc|}
        \hline
        \multirow{2}{*}{Dataset} & \multicolumn{3}{c|}{Training Stages} & \multicolumn{2}{c|}{AUROC} \\
        %\cline{2-6}
         &  \stageone{} &  \stagetwo{} &  \stagethree{} & Image-level & Pixel-level  \\
          \hline
          \hline
          \multirow{4}{*}{BraTS2021} 
          & \textcolor{ForestGreen}{\checkmark} & \textcolor{Red}{\xmark} & \textcolor{Red}{\xmark} & $89.3$ & $96.9$ \\
          & \textcolor{Red}{\xmark} & \textcolor{ForestGreen}{\checkmark}  & \textcolor{Red}{\xmark} & $82.3$ & $95.5$ \\
          & \textcolor{Red}{\xmark}  & \textcolor{Red}{\xmark} & \textcolor{ForestGreen}{\checkmark} & $90.1$ & -  \\
          & \textcolor{ForestGreen}{\checkmark}  & \textcolor{ForestGreen}{\checkmark} & \textcolor{ForestGreen}{\checkmark} & $94.1$ & $97.3$ \\
          \hline
          \multirow{4}{*}{BTCV + LiTS} 
          & \textcolor{ForestGreen}{\checkmark} & \textcolor{Red}{\xmark} & \textcolor{Red}{\xmark} & $83.7$ & $99.6$ \\
          & \textcolor{Red}{\xmark} & \textcolor{ForestGreen}{\checkmark}  & \textcolor{Red}{\xmark} & $63.0$ & $99.4$ \\
          & \textcolor{Red}{\xmark}  & \textcolor{Red}{\xmark} & \textcolor{ForestGreen}{\checkmark} & $71.0$ & -  \\
          & \textcolor{ForestGreen}{\checkmark}  & \textcolor{ForestGreen}{\checkmark} & \textcolor{ForestGreen}{\checkmark} & $84.2$ & $99.6$  \\
          \hline
          \multirow{4}{*}{RESC} 
          & \textcolor{ForestGreen}{\checkmark} & \textcolor{Red}{\xmark} & \textcolor{Red}{\xmark} & $65.0$ & $86.4$ \\
          & \textcolor{Red}{\xmark} & \textcolor{ForestGreen}{\checkmark}  & \textcolor{Red}{\xmark} & $85.5$ & $95.6$ \\
          & \textcolor{Red}{\xmark}  & \textcolor{Red}{\xmark} & \textcolor{ForestGreen}{\checkmark} & $82.1$ & - \\
          & \textcolor{ForestGreen}{\checkmark}  & \textcolor{ForestGreen}{\checkmark} & \textcolor{ForestGreen}{\checkmark} & $88.3$ & $95.1$  \\
          \hline
          \multirow{4}{*}{OCT2017} 
          & \textcolor{ForestGreen}{\checkmark} & \textcolor{Red}{\xmark} & \textcolor{Red}{\xmark} & $71.1$ & - \\
          & \textcolor{Red}{\xmark} & \textcolor{ForestGreen}{\checkmark}  & \textcolor{Red}{\xmark} & $89.3$ & - \\
          & \textcolor{Red}{\xmark}  & \textcolor{Red}{\xmark} & \textcolor{ForestGreen}{\checkmark} & $87.3$ & - \\
          & \textcolor{ForestGreen}{\checkmark}  & \textcolor{ForestGreen}{\checkmark} & \textcolor{ForestGreen}{\checkmark} & $96.8$ & - \\
          \hline
          \multirow{4}{*}{RSNA} 
          & \textcolor{ForestGreen}{\checkmark} & \textcolor{Red}{\xmark} & \textcolor{Red}{\xmark} & $81.4$ & -  \\
          & \textcolor{Red}{\xmark} & \textcolor{ForestGreen}{\checkmark}  & \textcolor{Red}{\xmark} & $83.5$ & -   \\
          & \textcolor{Red}{\xmark}  & \textcolor{Red}{\xmark} & \textcolor{ForestGreen}{\checkmark} & $69.4$ & -   \\
          & \textcolor{ForestGreen}{\checkmark}  & \textcolor{ForestGreen}{\checkmark} & \textcolor{ForestGreen}{\checkmark} & $84.6$ & -   \\
          \hline
          \multirow{4}{*}{CAMELYON16} 
          & \textcolor{ForestGreen}{\checkmark} & \textcolor{Red}{\xmark} & \textcolor{Red}{\xmark} & $72.5$ & -  \\
          & \textcolor{Red}{\xmark} & \textcolor{ForestGreen}{\checkmark}  & \textcolor{Red}{\xmark} & $80.7$ & -   \\
          & \textcolor{Red}{\xmark}  & \textcolor{Red}{\xmark} & \textcolor{ForestGreen}{\checkmark} & $68.6$ & -   \\
          & \textcolor{ForestGreen}{\checkmark}  & \textcolor{ForestGreen}{\checkmark} & \textcolor{ForestGreen}{\checkmark} & $82.0$ & -   \\
    \hline
    \end{tabular}
    }
    \vspace{-0.25cm}
    \caption{Ablation study on the impact of each MADJEPA stage on the final model performance, measured in terms of both image-level AUROC and pixel-level AUROC (where possible).}
    \label{tab:ablation_stages}
    \vspace{-0.4cm}
\end{table}

\noindent \textbf{Ablation studies}. We first examine the contribution of each individual MADJEPA stage to the overall system performance. This ablation study is conducted across all six datasets, with results presented in Table \ref{tab:ablation_stages}. Across all datasets, aggregating the anomaly scores from the three stages yields higher overall performance than any individual stage alone. The most notable example is the OCT2017 dataset, where none of the individual stages exceed $90\%$ AUROC in isolation, while aggregating the scores from the three stages allows the system to attain $96.8\%$ AUROC. 

A further observation is that the individual stage achieving the highest performance varies across datasets. Specifically, the I-JEPA stage attains the highest standalone performance on BTCV+LiTS, the distillation stage achieves the best individual results on RESC, OCT2017, RSNA and CAMELYON16, while the pseudo-anomaly classifier performs the best on BraTS2021.

An additional noteworthy finding is that, on the three datasets where MADJEPA achieves state-of-the-art performance, namely BTCV+LiTS, RSNA and CAMELYON16, our method preserves the same or a similar ranking even when relying on the anomaly scores of a single stage alone. Specifically, the I-JEPA stage attains state-of-the-art performance on BTCV+LiTS with $83.7\%$ AUROC, the distilled model reaches state-of-the-art performance on CAMELYON with $80.7\%$ AUROC, and the distilled model attains near state-of-the-art performance on RSNA with $83.5\%$ AUROC.

\begin{figure}[t]
    \centering
    \includegraphics[width=0.62\linewidth]{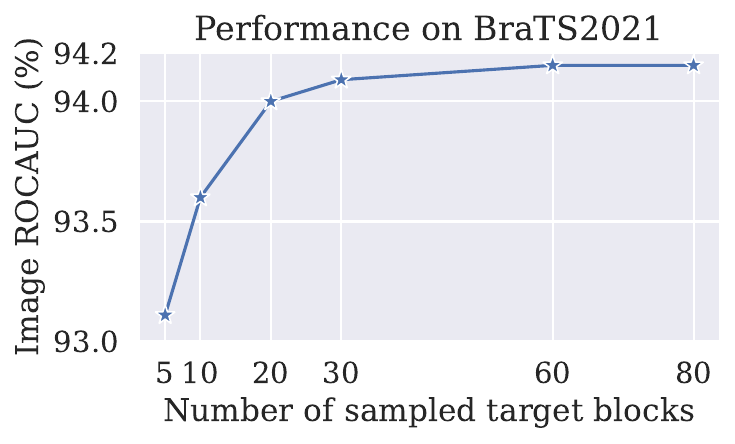}
    \hfill
    \includegraphics[width=0.62\linewidth]{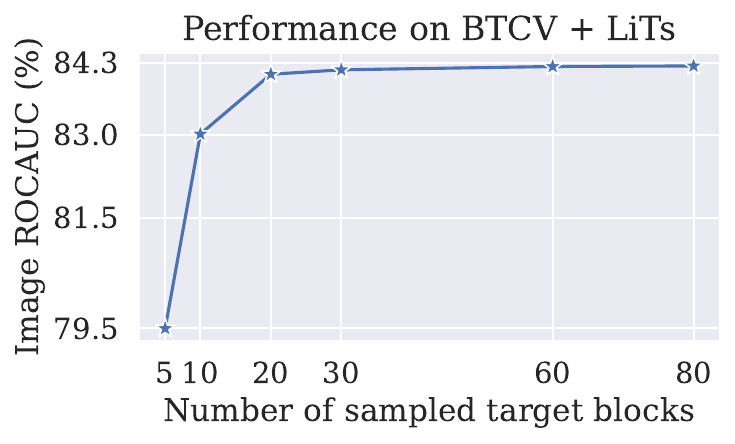}
    \vspace{-0.35cm}
    \caption{Ablation on the number of sampled target blocks ($N$) on BraTS2021 (top) and BTCV+LiTS (bottom). The performance of MADJEPA increases with the number of target blocks. However, on both datasets, beyond $30$ sampled blocks, the performance gain is marginal.}
    \label{fig:ablation_target_blocks}
     \vspace{-0.35cm}
\end{figure}

Furthermore, we analyze the impact of the number of sampled target blocks ($N$). During inference, we sequentially sample multiple target blocks and aggregate the per-patch distances between the target encoder and the predictor representations. For an anomalous image, sampling a larger number of blocks increases the likelihood of capturing patches belonging to the anomalous region within multiple target boxes. However, this comes at the cost of increased inference time. To characterize this trade-off, we examine how model performance varies with the number of sampled target blocks on two of our datasets, BraTS2021 and BTCV+LiTS, illustrated in Figure~\ref{fig:ablation_target_blocks}. On both datasets, we observe that performance increases only marginally beyond $30$ sampled blocks, such that the resulting increase in latency is not justified by the modest performance gain. We therefore set the number of target blocks $N$ to $30$ in all experiments.

\vspace{-0.1cm}
\section{Conclusion}
\label{ref:conclusion}
\vspace{-0.1cm}
In this work, we introduced MADJEPA, a multi-stage joint-embedding predictive architecture for unsupervised medical anomaly detection. %Rather than the usual pixel-level reconstruction objective, our approach learns contextual representations of normal anatomy in latent space. 
Our ablation results showed that I-JEPA alone is insufficient for anomaly detection, as its representations can either over-penalize unseen healthy structures or become too invariant to subtle pathologies. This is mostly due to the open-set nature of medical anomaly detection. MADJEPA addresses this issue with three complementary training stages: firstly, I-JEPA establishes a semantic representation of the normal anatomy, secondly, teacher-student distillation then introduces a discrepancy signal that curbs over generalization to abnormal structures, and thirdly, pseudo-anomaly training calibrates the decision boundary around the normal manifold. Together, these stages turn a general-purpose predictive world model into a representation specifically calibrated for medical anomaly detection.

Across all six BMAD datasets, spanning MRI, CT, OCT, X-ray, and histopathology, MADJEPA performs competitively against state-of-the-art methods, being the only method surpassing the $80\%$ threshold on all six datasets. % The complementary ablation studies confirm that each stage has a positive contribution. 
More broadly, these results suggest that predictive world models stand as a promising foundation for medical anomaly detection. We hope this motivates further work on such models as an alternative to pixel-level reconstruction and conventional self-supervised objectives.

\vspace{-0.1cm}
\section*{Acknowledgment}
\vspace{-0.1cm}

This work was supported by a grant of the Ministry of Research, Innovation and Digitization, CCCDI - UEFISCDI, project number PN-IV-P7-7.1-PED-2024-1856, within PNCDI IV.

{
    \small
    \bibliographystyle{ieeenat_fullname}
    \bibliography{refs}
}

\clearpage

\setcounter{section}{0}
\renewcommand{\thesection}{S\arabic{section}}

\setcounter{table}{0}
\renewcommand{\thetable}{S\arabic{table}}

\setcounter{figure}{0}
\renewcommand{\thefigure}{S\arabic{figure}}

\section*{\Large{Supplementary Material}}

% \vspace{-0.3cm}
% \section{Data preprocessing}
% \label{sec:data_prep}
% \vspace{-0.1cm}

% \textbf{Region of Interest (ROI) cropping}

% The images in four of the datasets from the BMAD benchmark (BraTS2021, BTCV+LiTs, RESC, OCT2017) contain a substantial area of background, which hinders the ability of I-JEPA to correctly learn the latent representation of the normal tissue. Due to the inherent random nature of the target block selection, the model learns to predict the latent representation of the background patches for a large proportion of target blocks. This slows the learning process, as the model must be trained on a larger number of target blocks to be exposed to a sufficient amount of information from the normal tissue. Consequently, for the four datasets, we introduce an additional preprocessing step consisting of region-of-interest (ROI) identification followed by ROI cropping for enabling the model to focus on the most informative region of the image. We first identify the bounding box that fully encloses the organ tissue, treating the zero-valued pixels as background and subsequently crop the image according to the coordinates of this bounding box. This preprocessing step is applied across all three training phases: I-JEPA training (Stage \stageone{}), target encoder distillation (Stage \stagetwo{}) and pseudo-anomaly classifier training (Stage \stagethree{}).

\section{Additional Ablations}
\label{sec:additional_ablations}

\textbf{Impact of patch-level pooling}. In Table~\ref{tab:ablation_patch_agg}, we investigate the effect of the patch-level pooling strategy used when building the anomaly maps. We evaluate it on the BraTS2021 and CAMELYON16 test splits. During stage \stageone{} inference, multiple blocks are sampled and patches contained within more than one target block are consequently assigned multiple distance values. Therefore, we experiment with both mean and maximum pooling functions for each dataset. %, selecting the strategy that yields the best performance on the official validation split. 
On BraTS2021, max pooling achieves a marginally higher score than our selected configuration (shown in bold) on the test set ($94.4\%$ vs.~$94.1\%$). However, this selection was made based on validation set performance, and the test set outcome was not observed at selection time. On CAMELYON16, mean pooling, the selected strategy, yields a better result than max pooling on the test set ($82.0\%$ vs.~$81.9\%$), consistent with the selection made on the validation split. In both cases, the differences are rather marginal.

\begin{table}[t]
    \centering
    \setlength\tabcolsep{0.06cm}
    %\resizebox{0.99\linewidth}{!}
    \fontsize{8}{9}\selectfont{
    \begin{tabular}{|l|c|c|c|}
        \hline
        Dataset & \makecell{Patch-level \\ pooling} & \makecell{Image-level \\ AUROC} & \makecell{Pixel-level \\ AUROC} \\
          \hline
          \hline
          \multirow{2}{*}{BraTS2021} 
           & \textbf{mean} & \textbf{94.1}  & \textbf{97.3}\\
           & max & 94.4  & 97.4  \\

          \hline
          \multirow{2}{*}{CAMELYON16} 
           & \textbf{mean} & \textbf{82.0} & - \\
           & {max} & {81.9} & {-} \\
           \hline
           
    \end{tabular}
    }
     \vspace{-0.2cm}
    \caption{Ablation study on patch-level pooling. We report image-level and pixel-level AUROC on BraTS2021, and image-level AUROC on CAMELYON16. The configuration used to produce our final results, selected on the official validation split, is indicated in bold.}
    \label{tab:ablation_patch_agg}
 \vspace{-0.2cm}
\end{table}

\noindent
\textbf{Impact of global pooling.} In Table~\ref{tab:ablation_image_agg}, we examine the influence of the global (image-level) pooling of the anomaly maps, on BraTS2021 and CAMELYON16 test sets. Specifically, after constructing an anomaly map, we aggregate the pixel values into a unique score employing one of the two pooling functions: mean or maximum. 
On BraTS2021, max pooling over the anomaly map yields the best result ($94.1\%$ vs.~$91.4\%$), with a non-negligible margin over mean pooling. On CAMELYON16, this effect is even more pronounced, with mean pooling substantially outperforming maximum pooling ($82.0\%$ vs.~$78.3\%$). On BraTS2021, since this aggregation choice affects only the image-level score and does not alter the underlying anomaly maps, the pixel-level AUROC remains unaffected across the two aggregation strategies. 

\begin{table}[t]
    \centering
    \setlength\tabcolsep{0.06cm}
    %\resizebox{0.99\linewidth}{!}
    \fontsize{8}{9}\selectfont{
    \begin{tabular}{|l|c|c|c|}
        \hline
        Dataset & \makecell{Image-level \\ pooling} & \makecell{Image-level \\ AUROC} & \makecell{Pixel-level \\ AUROC} \\
          \hline
          \hline
          \multirow{2}{*}{BraTS2021} 
           & mean & 91.4  & 97.3\\
           & \textbf{max} & \textbf{94.1}  & \textbf{97.3}  \\

          \hline
          \multirow{2}{*}{CAMELYON16} 
           & \textbf{mean} & \textbf{82.0} & - \\
           & max & {78.3} & {-} \\
           \hline
           
    \end{tabular}
    }
    \vspace{-0.2cm}
    \caption{Ablation study on image-level pooling. We report image-level and pixel-level AUROC on BraTS2021, and image-level AUROC on CAMELYON16. The configuration used to produce our final results, selected on the official validation split, is indicated in bold.}
    \label{tab:ablation_image_agg}
\vspace{-0.05cm}
\end{table}

\noindent
\textbf{Impact of the distance metric between patch embeddings.} In Table~\ref{tab:ablation_distance}, we explore the impact of the distance metric used for measuring the discrepancies between patch representations. Choosing the $L_2$ distance consistently yields the best results on both BraTS2021 and CAMELYON16, in terms of image-level AUROC. For BraTS2021, the findings remain consistent on the pixel-level metric. 

\begin{table}[t]
    \centering
    \setlength\tabcolsep{0.06cm}
    %\resizebox{0.99\linewidth}{!}
    \fontsize{8}{9}\selectfont{
    \begin{tabular}{|l|c|c|c|}
        \hline
        Dataset & \makecell{Distance\\ measure} & \makecell{Image-level \\ AUROC} & \makecell{Pixel-level \\ AUROC} \\
          \hline
          \hline
          \multirow{2}{*}{BraTS2021} 
           & $L_1$ & 94.1  &  96.9 \\
           & $\mathbf{L_2}$ & \textbf{94.1}  & \textbf{97.3}  \\

          \hline
          \multirow{2}{*}{CAMELYON16} 
           & $L_1$  & 81.9 & - \\
           & $\mathbf{L_2}$ & \textbf{82.0} & \textbf{-} \\
           \hline
           
    \end{tabular}
    }
    \vspace{-0.2cm}
    \caption{Ablation study on the distance between patch-level representations. We report image-level and pixel-level AUROC on BraTS2021, and image-level AUROC on CAMELYON16. The configuration used to produce our final results, selected on the official validation split, is indicated in bold.}
    \label{tab:ablation_distance}
\vspace{-0.2cm}
\end{table}

\begin{figure}[t]
    \centering
    \includegraphics[width=0.65\linewidth]{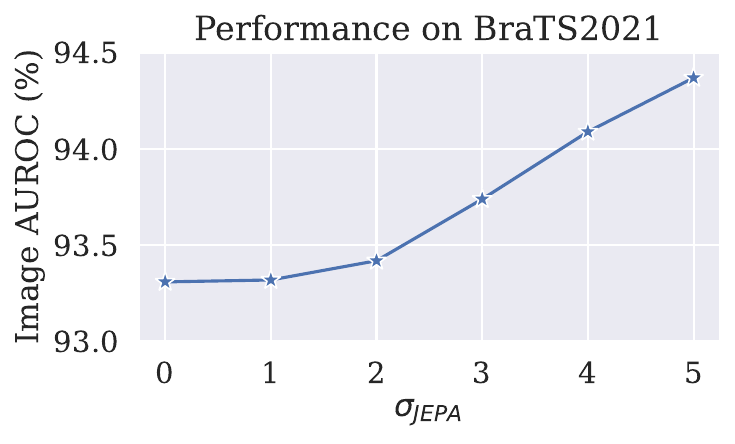}
    \hfill
    \includegraphics[width=0.65\linewidth]{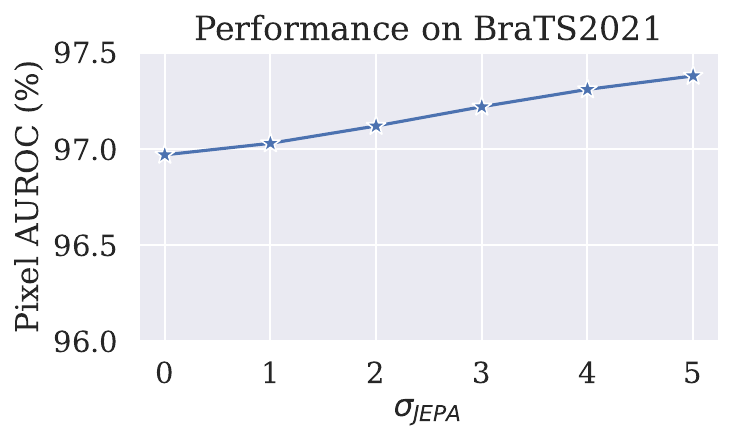}
    \vspace{-0.35cm}
    \caption{Ablation study on the Gaussian blurring factor $\sigma_\text{JEPA}$ for the anomaly maps generated in stage \stageone, evaluated on the BraTS2021 test set. We report the image-level AUROC (top) and pixel-level AUROC (bottom). The performance of MADJEPA increases with the value of $\sigma_\text{JEPA}$. However, on the validation set, gains beyond $\sigma_\text{JEPA} > 4$ are marginal. Consequently, we employ  $\sigma_\text{JEPA} = 4$ for the final results on BraTS2021.}
    \label{fig:ablation_sigma_jepa}
     \vspace{-0.3cm}
\end{figure}

\begin{figure}[t]
    \centering
    \includegraphics[width=0.65\linewidth]{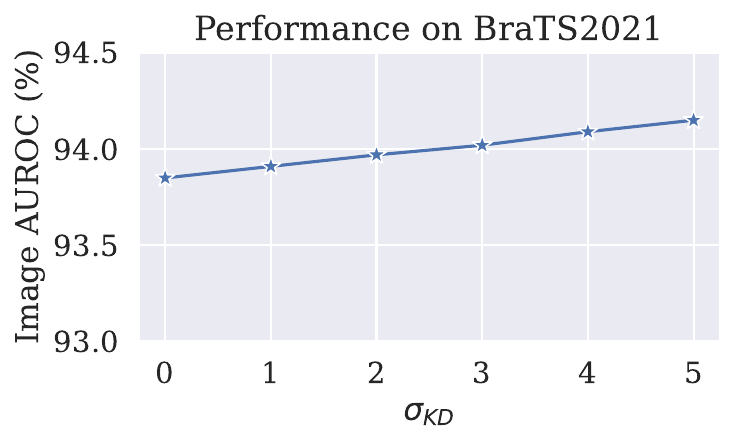}
    \hfill
    \includegraphics[width=0.65\linewidth]{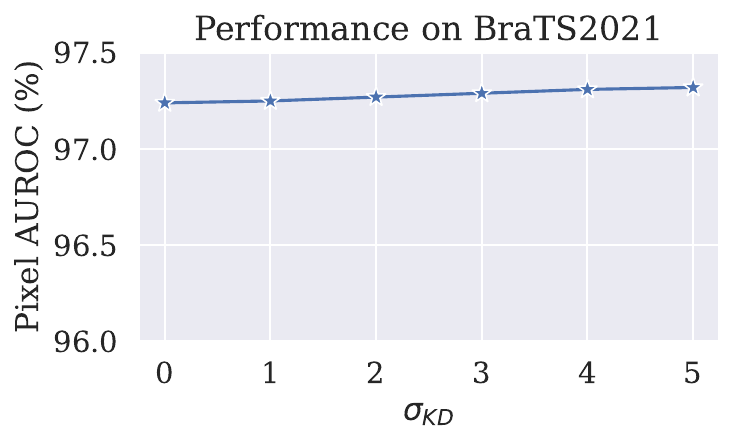}
    \vspace{-0.35cm}
    \caption{Ablation study on the Gaussian blurring factor $\sigma_\text{KD}$ for the anomaly maps generated in stage \stagetwo, evaluated on the BraTS2021 test set. We report the image-level AUROC (top) and pixel-level AUROC (bottom). The performance of MADJEPA is only marginally affected by the variation of $\sigma_\text{KD}$. Given these diminishing gains, we set $\sigma_\text{KD} = \sigma_\text{JEPA} = 4$ for all the experiments performed on BraTS2021.}
    \label{fig:ablation_sigma_distill}
     \vspace{-0.4cm}
\end{figure}

\begin{figure}[t]
    \centering
    \includegraphics[width=0.65\linewidth]{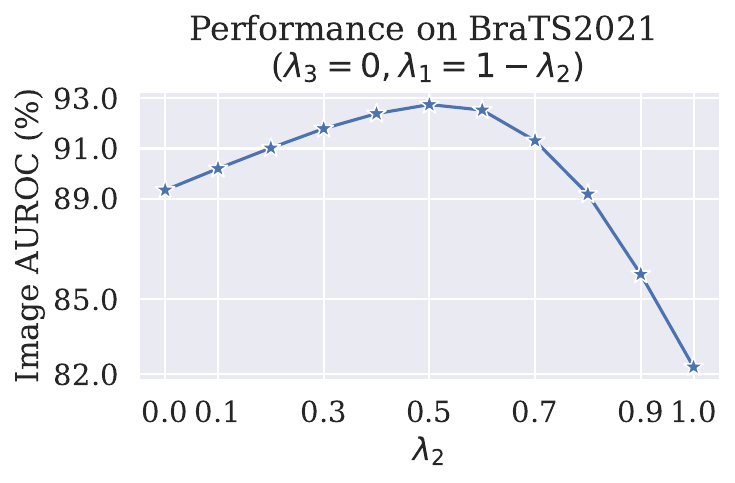}
    \hfill
    \includegraphics[width=0.65\linewidth]{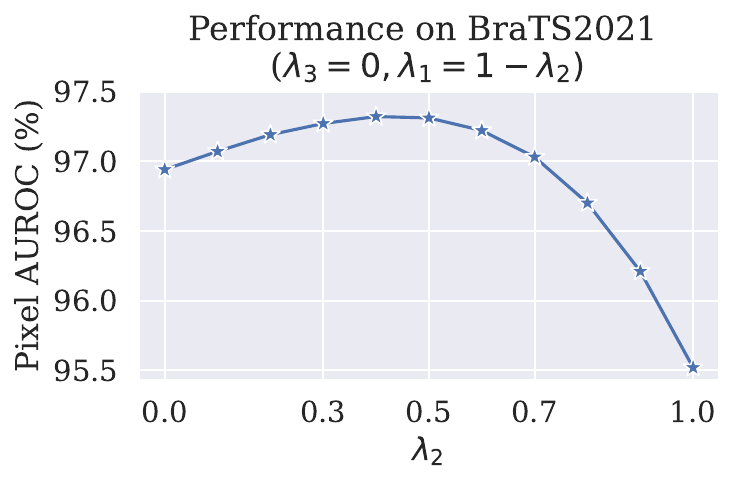}
    \vspace{-0.35cm}
    \caption{Ablation study on the anomaly score weights $\lambda_1$ and $\lambda_2$. We report image-level AUROC (top) and pixel-level AUROC (bottom) on BraTS2021 test set. We fix $\lambda_3 = 0$ and, since we impose the weights to sum to $1$, $\lambda_1 = 1 - \lambda_2$. The findings show that the best performance is obtained when $\lambda_1 = \lambda_2 = 0.5$, which is consistent with the selection we made on the official validation set. Specifically, for all the experiments performed on BraTS2021, we employ equal weights for stage \stageone{} and stage \stagetwo{} anomaly scores.}
    \label{fig:ablation_lambda_2}
     \vspace{-0.3cm}
\end{figure}

\begin{figure}[t]
    \centering
    \includegraphics[width=0.65\linewidth]{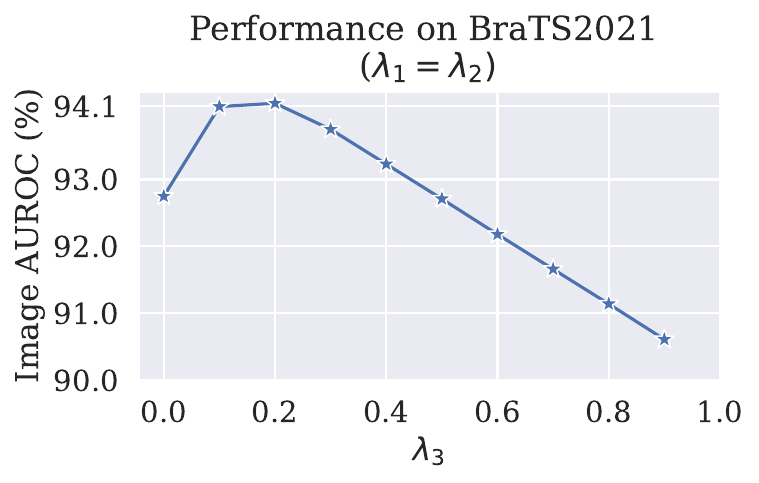}
    \hfill
    \includegraphics[width=0.65\linewidth]{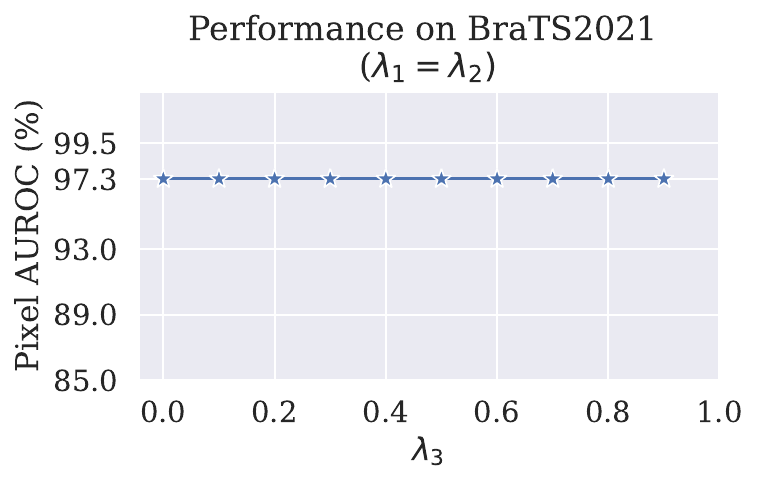}
    \vspace{-0.35cm}
    \caption{Ablation study on the anomaly score weight $\lambda_3$. We report image-level AUROC (top) and pixel-level AUROC (bottom) on BraTS2021. For this ablation, we fix $\lambda_1 = \lambda_2$. Setting $\lambda_3 = 0.1$ or $\lambda_3 = 0.2$ yields the highest image-level AUROC values on the test set. The classification head does not influence the anomaly maps, so changing $\lambda_3$ does not modify the pixel-level AUROC.}
    \label{fig:ablation_lambda_3}
     \vspace{-0.4cm}
\end{figure}

\noindent
\textbf{Impact of the Gaussian blurring factors.} As illustrated in Figure~\ref{fig:ablation_sigma_jepa}, the performance of MADJEPA is considerably influenced by the value of $\sigma_\text{JEPA}$. In terms of image-level AUROC, the performance increases from $93.3\%$ without Gaussian blurring ($\sigma_\text{JEPA} = 0$) to $94.4\%$ when $\sigma_\text{JEPA} = 5$. Regarding the pixel-level AUROC, the gains are less pronounced, increasing from $97.0\%$ ($\sigma_\text{JEPA} = 0$) to $97.4\%$ ($\sigma_\text{JEPA} = 5$). These results highlight that the performance of MADJEPA increases with $\sigma_\text{JEPA}$. However, on the validation set, gains beyond $\sigma_\text{JEPA} > 4$ are marginal. Therefore, we set $\sigma_\text{JEPA} = 4$ for the final results reported on BraTS2021.

Conversely, in Figure~\ref{fig:ablation_sigma_distill}, we notice that the performance of MADJEPA is marginally influenced by the value of $\sigma_\text{KD}$. Considering that the gains are reduced, for convenience, we set $\sigma_\text{KD}$ to be equal to $\sigma_\text{JEPA}$. Consequently, we employed $\sigma_\text{KD} = 4$ for all the experiments we conducted on BraTS2021.

\begin{table*}[t]
    \centering
    \setlength\tabcolsep{0.06cm}
    %\resizebox{0.99\linewidth}{!}
    \fontsize{8}{9}\selectfont{
    \begin{tabular}{| l | l | ccc | ccc | ccc |}
        \hline
        \multirow{2}{*}{Dataset} & \multirow{2}{*}{Medical domain} & \multicolumn{3}{c|}{Train set} & \multicolumn{3}{c|}{Validation set} & \multicolumn{3}{c|}{Test set} \\
        \cline{3-11}
         & & Normal & Anomaly & Image size & Normal & Anomaly & Image size & Normal & Anomaly & Image size \\ 
        \hline
        \hline
          BraTS2021 & Brain MRI & 7500  & 0  & 240$\times$240 & 39 & 44 & 240$\times$240 & 640 & 3075 & 240$\times$240 \\
        \hline
          BTCV+LiTS & Liver CT & 1542  & 0  & 512$\times$512 & 93 & 73 & 512$\times$512 & 833 & 660 & 512$\times$512 \\
        \hline
          RESC & Retinal OCT  & 4297 & 0 & 1024$\times$512  & 45 & 70 & 1024$\times$512 & 1041 & 764 & 1024$\times$512 \\
        \hline
          OCT2017 & Retinal OCT & 26315 & 0 & \makecell{496-512$\times$384-1536}  & 8 & 24 &  \makecell{496$\times$512/768} & 242 & 726 & \makecell{496$\times$512-1536} \\
        \hline
          RSNA & Chest X-ray  & 8000 & 0 & 1024$\times$1024  & 70 & 1420 & 1024$\times$1024 & 781 & 16413 & 1024$\times$1024 \\
        \hline
          CAMELYON16 & Digital Histopathology  & 5088 & 0 & 256$\times$256  & 117 & 119 & 256$\times$256 & 1003 & 994 & 256$\times$256 \\
    \hline
    \end{tabular}}
    \vspace{-0.2cm}
    \caption{Overview of the six medical imaging datasets included in BMAD benchmark.}
    \label{tab:dataset}
    \vspace{-0.1cm}
\end{table*}

\begin{figure*}
    \centering
    \includegraphics[width=0.96\linewidth]{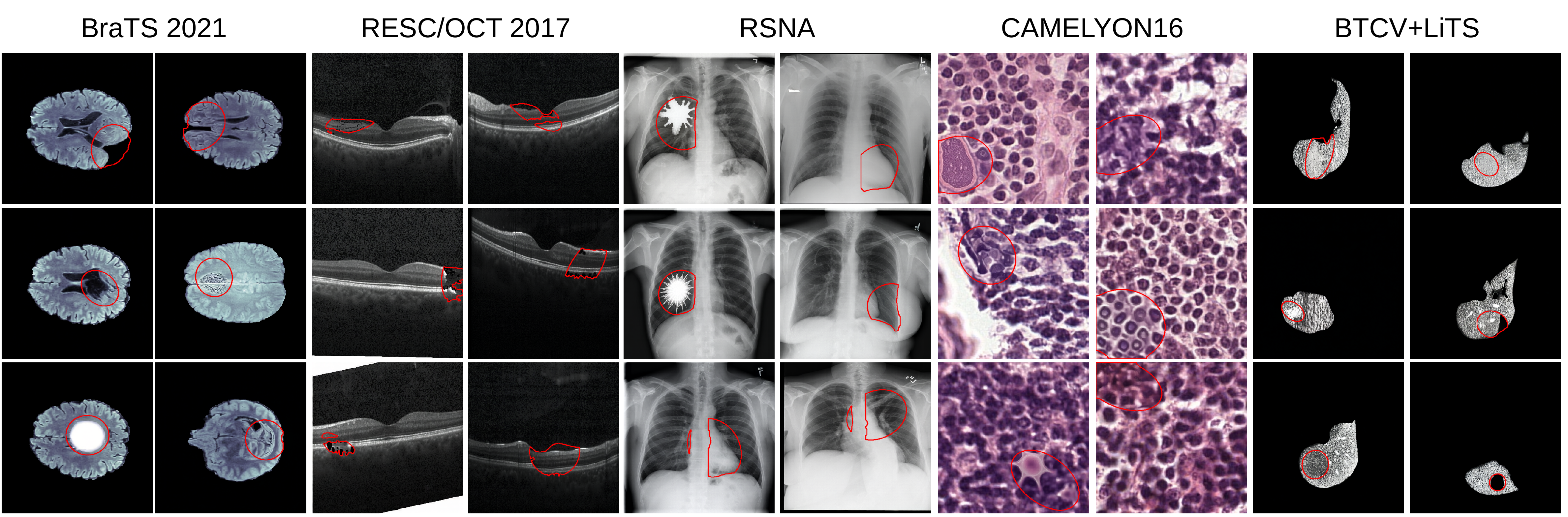}
    \vspace{-0.3cm}
    \caption{Examples of synthetic pseudo-anomalies generated with Qwen-Image~\cite{Wu-arXiv-2025}. These are used in training stage \stagethree{} of MADJEPA. Best viewed in color.}
    \label{fig:anomalies}
    \vspace{-0.3cm}
\end{figure*}

\noindent \textbf{Impact of anomaly score weights.} We next examine the influence of the anomaly score weights, namely $\lambda_1$, $\lambda_2$ and $\lambda_3$, on the BraTS2021 dataset. We first fix $\lambda_3 = 0$ and search for the optimal weighting between $\lambda_1$ and $\lambda_2$. As shown in Figure~\ref{fig:ablation_lambda_2}, the setting $\lambda_1 = \lambda_2 = 0.5$ yields the best performance. This observation aligns with the coefficients selected on the official validation split, where we set $\lambda_1$ = $\lambda_2 = 0.45$ for the final results on BraTS2021. 

Second, we fix the ratio between the first two coefficients, as identified in the previous step (in our case, $\lambda_1 = \lambda_2$), and we search for the optimal value of the third coefficient, $\lambda_3$. As illustrated in Figure~\ref{fig:ablation_lambda_3}, setting $\lambda_3 = 0.1$ or $\lambda_3 = 0.2$ yields the largest image-level AUROC. In terms of pixel-level AUROC, the choice of $\lambda_3$ does not have any impact. This result is consistent with the anomaly score weights selected on the validation set, namely: $\lambda_1=0.45$, $\lambda_2=0.45$, $\lambda_3=0.1$. Interestingly, by further examining Figure~\ref{fig:ablation_lambda_3}, we observe that for any value of $\lambda_3$, the AUROC score does not decrease under $90\%$. This highlights the robustness of our method to the choice of hyperparameters. While the values of the three coefficients matter, the performance remains competitive even when the values of the weights are potentially suboptimal.

\section{Data Details}
\label{sec:bmad}

\noindent \textbf{BMAD splits.}
In Table~\ref{tab:dataset}, we provide a detailed overview of the official training, validation, and test splits of the BMAD benchmark \cite{Bao-CVPRW-2024} for all six datasets, including the distribution of samples across each split.

\noindent \textbf{Synthetic pseudo-anomalies.}
In Figure~\ref{fig:anomalies}, we showcase some synthetic samples used in our pseudo-anomaly classifier. Starting from healthy samples, we apply inpainting inside the regions marked by the red contours, using Qwen-Image~\cite{Wu-arXiv-2025}.

\begin{table}[!th]
    \centering
    \setlength\tabcolsep{0.06cm}
    %\resizebox{0.99\linewidth}{!}
    \fontsize{8}{9}\selectfont{
        \begin{tabular}{|l|l|c|}
        \hline
        Dataset & Hyperparameter & Value \\
          \hline
          \hline
          \multirow{9}{*}{BraTS2021} 
           & Number of sampled target boxes ($N$) & 30  \\
           & Patch-level aggregation strategy & mean \\
           & Image-level aggregation strategy & max    \\
           & Distance for patch-level representations & $L_2$  \\
           & JEPA weight ($\lambda_1$) & 0.45 \\
           & KD weight ($\lambda_2$) & 0.45 \\
           & Classification weight ($\lambda_3$) & 0.10 \\
           & $\sigma_{\text{JEPA}}$ for Gaussian blur & 4 \\
           & $\sigma_{\text{KD}}$ for Gaussian blur & 4 \\
          \hline
          \multirow{9}{*}{BTCV+LiTS} 
           & Number of sampled target boxes ($N$) & 30  \\
           & Patch-level aggregation strategy & mean \\
           & Image-level aggregation strategy & max    \\
           & Distance for patch-level representations & $L_2$  \\
           & JEPA weight ($\lambda_1$) & 0.74 \\
           & KD weight ($\lambda_2$) & 0.24 \\
           & Classification weight ($\lambda_3$) & 0.02 \\
           & $\sigma_{\text{JEPA}}$ for Gaussian blur & 4 \\
           & $\sigma_{\text{KD}}$ for Gaussian blur & 4 \\
          \hline
          \multirow{9}{*}{RESC} 
           & Number of sampled target boxes ($N$) & 30  \\
           & Patch-level aggregation strategy & max \\
           & Image-level aggregation strategy & max    \\
           & Distance for patch-level representations & $L_2$  \\
           & JEPA weight ($\lambda_1$) & 0.23 \\
           & KD weight ($\lambda_2$) & 0.52 \\
           & Classification weight ($\lambda_3$) & 0.25 \\
           & $\sigma_{\text{JEPA}}$ for Gaussian blur & 4 \\
           & $\sigma_{\text{KD}}$ for Gaussian blur & 2 \\
          \hline
          \multirow{9}{*}{OCT2017} 
           & Number of sampled target boxes ($N$)  & 30  \\
           & Patch-level aggregation strategy & max \\
           & Image-level aggregation strategy & max    \\
           & Distance for patch-level representations & $L_1$  \\
           & JEPA weight ($\lambda_1$) & 0.45 \\
           & KD weight ($\lambda_2$) & 0.45 \\
           & Classification weight ($\lambda_3$) & 0.10 \\
           & $\sigma_{\text{JEPA}}$ for Gaussian blur & 1 \\
           & $\sigma_{\text{KD}}$ for Gaussian blur & 1 \\
          \hline
          \multirow{9}{*}{RSNA} 
           & Number of sampled target boxes ($N$) & 30  \\
           & Patch-level aggregation strategy & max \\
           & Image-level aggregation strategy & mean    \\
           & Distance for patch-level representations & $L_2$  \\
           & JEPA weight ($\lambda_1$) & 0.29 \\
           & KD weight ($\lambda_2$) & 0.69 \\
           & Classification weight ($\lambda_3$)& 0.02 \\
           & $\sigma_{\text{JEPA}}$ for Gaussian blur & 4 \\
           & $\sigma_{\text{KD}}$ for Gaussian blur & 4 \\
          \hline
          \multirow{9}{*}{CAMELYON16} 
           & Number of sampled target boxes ($N$) & 30  \\
           & Patch-level aggregation strategy & mean \\
           & Image-level aggregation strategy & mean \\
           & Distance for patch-level representations & $L_2$  \\
           & JEPA weight ($\lambda_1$) & 0.10 \\
           & KD weight ($\lambda_2$) & 0.80 \\
           & Classification weight ($\lambda_3$)& 0.10 \\
           & $\sigma_{\text{JEPA}}$ for Gaussian blur & 4 \\
           & $\sigma_{\text{KD}}$ for Gaussian blur & 4 \\
    \hline
    \end{tabular}
    }
    \vspace{-0.2cm}
    \caption{{Inference hyperparameters for our anomaly detection pipeline}. The values of the hyperparameters are selected using grid search on the official validation splits.}
    \label{tab:inference_hyperparameters}
 %\vspace{-0.2cm}
\end{table}

\section{Hyperparameter Configurations}
\label{sec:hyperparams}

In Table~\ref{tab:inference_hyperparameters}, we provide a comprehensive list of all hyperparameters used by MADJEPA, such as the task weights or the Gaussian smoothing factor $\sigma$.

\section{Training and inference details}
\textbf{Compute environment.}
All experiments are performed across two workstations, each equipped with an NVIDIA GeForce RTX 3090 GPU (24GB VRAM), on a system leveraging 64GB of RAM, and CPUs with 24 and 12 cores, respectively. Each individual experiment ran on a single GPU, without model or data parallelism. 

\noindent
\textbf{Training time for each stage.} 
In Table~\ref{tab:training_time}, we report the training time (in hours) per stage for each dataset. On most datasets, the total training time, accumulated over the three stages, is under five hours. The only exception is OCT2017, the largest dataset from BMAD, which requires approximately seven hours of training. On average, the most demanding stage is represented by the target encoder distillation. This stage usually requires more training epochs due to the student model, which is randomly initialized and trained from scratch. 

\noindent
\textbf{Inference time details.} In stage \stageone{}, the inference time is monotonically increasing with the number of sampled blocks ($N$). For $N=30$, the average inference time per image is $0.8858$ seconds, with a standard deviation of $0.0071$ seconds.

% For each image, the target encoder is run only once in order to produce the patch representations for the entire input, requiring $T_t$ seconds for a single pass. However, we need to run the context encoder and the predictor for each sampled context and random block respectively. Let $T_c$ and $T_p$ be the time required by the context encoder and the predictor, respectively, for a single block. Therefore, for each image, the inference time could be computed as: $T_t + N \cdot (T_c + T_p)$. From our tests, the average inference time per image for stage \stageone{} is $0.8858$ (s) with a standard deviation of $0.0071$ (s) computed across five runs.

\noindent
\textbf{}

\begin{table}[!t]
    \centering
    \setlength\tabcolsep{0.06cm}
    %\resizebox{0.99\linewidth}{!}
    \fontsize{8}{9}\selectfont{
        \begin{tabular}{|l|c|c|c|}
        \hline
        Dataset & Stage \stageone & Stage \stagetwo & Stage \stagethree \\
          \hline
          \hline
          \multirow{1}{*}{BraTS2021} 
          & $<1$ h & $\approx2$ h &  $<1$ h  \\
          \hline

          \multirow{1}{*}{BTCV+LiTS} 
          & $<1$ h & $<1$ h & $<1$ h  \\
          \hline

          \multirow{1}{*}{RESC} 
          & $<1$ h & $\approx1$h & $<1$ h  \\
          \hline

          \multirow{1}{*}{OCT2017} 
          & $\approx2$ h & $\approx4$ h & $\approx1$ h  \\
          \hline

          \multirow{1}{*}{RSNA} 
          & $<1$h & $\approx3$ h & $<1$ h  \\
          \hline

          \multirow{1}{*}{CAMELYON16} 
          & $<1$ h & $\approx1$ h & $<1$ h  \\
          \hline

    % \hline
    \end{tabular}
    }
    \vspace{-0.2cm}
    \caption{Training time (in hours) for each stage, reported per dataset.}
    \label{tab:training_time}
\vspace{-0.1cm}
\end{table}

\end{document}